\documentclass[12pt,aps]{revtex4}
\usepackage{epsfig,amsmath,amssymb}

\newcommand{\fr}[1]{%   arg1: var name
             \frac{#1}}
\newcommand{\bea}{\begin{eqnarray}}
\newcommand{\eea}{\end{eqnarray}}

\newcommand{\ket}{{\rangle}}

\newcommand{\bra}{{\langle}}
\newcommand{\gc}{\bra\fr{\alpha_s}{\pi}G^2\ket}

\newcommand{\ga}{g_{{\cal A}}}

\newcommand{\dbar}{d\hskip -0.4em ^-}

\newcommand{\mc}[1]{{\mathcal{#1}}}
\begin{document}

\title{Non-factorizable contribtion to 
  $\overline{B_{d}^0} \rightarrow \, \pi^0 D^{0}$} 

\vspace{0.5cm}

%\maketitle

\author{Lars E. Leganger}
%\email{l.e.leganger@gmail.com}
\affiliation{Department of Physics, NTNU, 
Trondheim, Norway}

\author{Jan O. Eeg}
%\email{j.o.eeg@fys.uio.no}
\affiliation{Department of Physics, University of Oslo,
P.O.Box 1048 Blindern, N-0316 Oslo, Norway}

\vspace{0.5 cm}

\begin{abstract}

The  decay modes of the type 
$B \rightarrow \pi \, D $ are dynamically different. 
For the case  $\overline{B_{d}^0} \rightarrow \, \pi^- D^{+} $
there is a substantial factorized contribution which dominates.
In contrast,  the decay mode 
  $\overline{B_{d}^0} \rightarrow \, \pi^0 D^{0} $ 
has a small  factorized contribution, being proportional to a very small Wilson
  coefficient combination. In this paper we calculate the relevant 
Wilson coefficients  at one loop level 
in the
 heavy  quark limits, both for the $b$-quark and the $c$-quark.

We also emphasize that
for  the decay mode
  $\overline{B_{d}^0} \rightarrow \, \pi^0 D^{0} $
there is a  sizeable  non-factorizable contribution due long distance 
interactions, which
dominate the amplitude. 
We estimate the branching ratio for this decay mode within our framework, 
which uses  the heavy quark limits, both for  the $b$- and the $c$-
quarks.
 In addition, we treat energetic light ($u,d,s$) quarks 
 within a variant of
 Large Energy Effective Theory  and combine this with a new extension
 of chiral quark models.

For reasonable values of the model dependent parameters of our model
can account for at least 3/4 of the   amplitude needed to explain
 the experimental branching ratio 
$ \simeq \, 2.6 \times 10^{-4}$.

\end{abstract}

\maketitle

\vspace{1cm}

Keywords:
%\begin{keyword}
$B$-decays, factorization, gluon condensate. \\
PACS:  13.20.Hw ,  12.39.St , 12.39.Fe ,  12.39.Hg.
%\end{keyword}

%\end{frontmatter}
%\vspace{1cm}

\newpage

\section{Introduction}

There is presently great interest in decays of $B$-mesons, 
due to numerous experimental results coming from BaBar and Belle. 
Soon  LHC will also provide data for such processes. 
$B$-decays of the type $B \rightarrow \pi \pi$ and $B \rightarrow K
\pi$,  
where  the energy
 release is big  compared to the light
meson masses,  has been treated within
{\it QCD factorization} \cite{BBNS} and
{\it Soft Collinear Effective  Theory} (SCET) \cite{SCET}. In 
 the high energy limit  
 the amplitudes for such decay modes factorize into  products
of two matrix elements of weak currents, 
and  some non-factorizable corrections of order $\alpha_s$
 which  can be calculated perturbatively.
 The decays $B \rightarrow \pi \pi,  K \pi$ are typical heavy to light
 decays.  
It was pointed out in previous papers \cite{EFHP} that 
 for various decays of the type $\bar{B} \rightarrow D \bar{D}$, which
 are of heavy to heavy type,
  the methods of \cite{BBNS,SCET} are not expected to hold because
  the energy release
is of order 1 GeV only. The 
 so-called pQCD model \cite{pQCD} was also 
used for such decay modes\cite{pQCDBDD}

The last two decades,   $b$-quarks, and some times 
also $c$-quarks,
were described within {\it Heavy Quark Effective Field Theory}
(HQEFT) \cite{neu}.
Some transitions of {\em heavy to heavy} type have, in the heavy quark 
limits $(1/m_b) \rightarrow 0$ {\em and } $(1/m_c) \rightarrow 0$, 
 been 
studied within
{\it Heavy Light Chiral Perturbation Theory} (HL$\chi$PT)
\cite{itchpt}. Typical cases are the 
Isgur-Wise function  for   $B \rightarrow D \;$ transition currents  \cite{IW},
 and $B - \bar{B}$ mixing \cite{ahjoeB}.
Also other $B \rightarrow D$ transitions, where the energy gap between
the initial ($B$-meson) state and the final state (including a $D$-meson)
are substantial,
 have been analyzed within such a framework \cite{IW,GriLe}, even if it
 is not ideal. Especially
 in cases where the 
factorized amplitude is almost zero, calculations of non-factorized
 amplitudes in terms of chiral loops or soft gluon emission estimated within
 a chiral quark model might be fruitful \cite{EHP,EFHP,MacDJoe},
 because they are expected to give results of
 reasonable order of magnitude.

The HQEFT covers processes where the heavy quarks carry the main
part of the momentum in each hadron.
 To describe processes where energetic light quarks emerge from decays
 of heavy $b$-quarks,  {\it Large Energy Effective  Theory} (LEET)
 was introduced \cite{GD} and used to study  
the current for $B \rightarrow \pi$ \cite{charles}.  
LEET does for energetic light quarks what  HQEFT does for heavy quarks.
In HQEFT one splits off the motion of the heavy quark from the 
 heavy quark field, 
thus obtaining a reduced field depending on the velocity of the heavy quark.
 In LEET  one splits off the large energy from the field of the
light  energetic quark,  thus obtaining an effective theory for a reduced 
energetic light quark field depending on a light-like four vector. 
 It was later shown that LEET in its
 initial formulation was incomplete and
 did not fully reproduce  infrared QCD physics
\cite{Uglea}. Then  LEET was further developed
 to include collinear gluons, and
became the Soft Collinear Effective Theory (SCET)  \cite{SCET}.

In this paper we consider  decay modes of the type 
$B \rightarrow  \pi \, D$.
 We restrict ourselves to
processes  where the $b$-quark decays. This means  
 the quark level  processes  
$b  \rightarrow  c d \bar{u} \, $.
Processes where the anti- $b$-quark decays proceed analogously.
The decay mode  $\overline{B_{d}^0} \rightarrow \, \pi^- \, D^{+}$
has a substantial factorized amplitude, given by the Isgur-Wise function
for $B \rightarrow D$ transition times the decay constant for $\pi^-$.
The relevant Wilson coefficient is also the maximum possible, namely of 
order one times
the  relevant Cabibbo-Kobayashi-Maskawa (CKM) quark mixing 
 factors and the Fermi coupling constant.
This is in contrast to the process 
$\overline{B_{d}^0} \rightarrow \, \pi^0 D^{0}$ which is color suppressed,
 as already  pointed out in ref. \cite{neupet}.

First we  point out that the factorized contribution to the decay mode
 $\overline{B_{d}^0} \rightarrow \, \pi^0 D^{0}$,
  given by
the $B \rightarrow \pi$ transition amplitude times the decay constant of the 
$D^0$ meson, is almost zero because it is proportional to  a very small Wilson
coefficient combination. In section III this combination will be calculated 
 explicitly at  one loop level completely within HQEFT, and scaled down to 
the scale $\mu \simeq \; $ 1 GeV where perturbative QCD is matched to our
long distance framework.

Second,  in the present paper we construct a modified  version of the
LEET used in \cite{charles} to study the 
$B \rightarrow \pi$ current,  and in the next step 
construct a new model which we call
{\it Large Energy Chiral Quark Model} (LE$\chi$QM) \cite{LELeg}.
 The mentioned incompleteness 
of LEET does not concern us here because 
we will combine LEET with chiral quark models ($\chi$QM)
\cite{chiqm,pider,BHitE}, containing only soft gluons making condensates. 
 In our model an energetic quark is bound to a soft quark
with an apriori unknown coupling,  as proposed in \cite{EHFiz}. The
unknown coupling is determined by calculating the  known 
$B \rightarrow \pi$ current matrix element within the model. This will fix 
  the unknown  coupling because the matrix element of this current is known
\cite{charles}.
   Then, in the next step, we use this coupling  
 to  calculate the non-factorized (color suppressed) 
amplitude contribution to  
 $\overline{B_{d}^0} \rightarrow \, \pi^0 D^{0}$ in terms of the lowest
 dimension gluon condensate, as have been done for other non-leptonic decays
\cite{ahjoe,ahjoeB,EHP,MacDJoe,BEF}. 
After the quarks have been
integrated out, we obtain an effective theory containing 
both soft light mesons as in HL$\chi$PT, and  also
 fields describing energetic light mesons. A similar idea with a 
combination of SCET  with 
HL$\chi$PT  is  considered in \cite{GrinPir}.
The LE$\chi$QM is constructed in analogy with the previous
 {\it Heavy Light Chiral Quark Model} (HL$\chi$QM)\cite{ahjoe}
 and may be considered to be an extension of that model.

In the next section (II) we present the
 weak four quark Lagrangian and its factorized and non-factorizable
 matrix elements.
In section III we calculate the Wilson coefficients at one loop level 
in the  heavy quark limits for both the $b$- and the $c$-quark.
 In section IV we present our version of LEET,
and in section V we present the new model LE$\chi$QM
to include energetic light quarks and mesons.
In section VI we  calculate the 
non-factorizable
matrix elements due to soft gluons expressed through the (model
dependent) quark condensate. 
 In section VII we give the results and 
conclusion.

\section{The effective Lagrangian at quark level}

We study $\bar{B}^0$ decays generated  by the weak quark process
 $b \rightarrow c\bar{u}d$ 
The  effective weak Lagrangian at quark level is \cite{QCDloop1,QCDloop2}
\begin{equation}
  {\cal L}_\mathrm{eff} = -
  \frac{G_F}{\sqrt{2}}V_\textrm{cb}V^*_\textrm{ud}\left[ c_A \, Q_A
 + c_B \,  Q_B \right],
\label{effLag}
\end{equation}
where the subscript $L$ denotes the left-handed fields: $q_L \equiv L \, q$,
 where $L \equiv (1\, - \, \gamma_5)/2$ is the left-handed projector 
in Dirac-space. 
The local operator products $Q_{A,B}$ are 
\begin{equation}
  Q_A = 4 \, \bar{c}_L \gamma_\mu  b_L \; \bar{d}_L \gamma^\mu  u_L
\quad ; \; \, 
  Q_B = 4 \, \bar{c}_L \gamma_\mu  u_L \;  \bar{d}_L \gamma^\mu  b_L .
\end{equation}
In these operators summation over color is implied. In (\ref{effLag}), 
 $c_A$ and $c_B$ are Wilson coefficients. At tree level $c_A=1$ and
$c_B=0$. At one loop level, a contribution to $c_B$ is also generated, and 
$c_A$ is slightly increased. These effects are handled in terms of 
the {\it Renormalization Group Equations} (RGE)\cite{QCDloop1,QCDloop2}.

Using the color matrix identity 
\begin{equation}
 2 \; t_{i n}^a \; t_{l j}^a \, = \, 
\delta_{i j}\delta_{l n}  \, - 
\,    \frac{1}{N_c} \delta_{i n} \delta_{l j} \; \; ,
% \; +  \; 2 \; t_{i n}^a \; t_{l j}^a  \; \, ,
\label{eq:color-identity}
\nonumber
\end{equation} 
 and Fierz rearrangement, 
the amplitudes for  decays of $\overline{B_d^0}$ into $D \, \pi$ 
may be written as
\begin{eqnarray}
% \begin{split}
   \mathcal{M}_{D^+\pi^-} \, =
\, 4 \, \frac{G_F}{\sqrt{2}}V_\textrm{cb}V^*_\textrm{ud}
 & \left[
     \left(c_A+\frac{1}{N_\textrm{c}}c_B \right)
     \bra{\pi^-} | \bar{d}_L \gamma_\mu \, u_L | 0 \ket \bra{D^+} | \bar{c}_L
\gamma^\mu \, b_L |\overline{B_d^0} \ket \right.   \nonumber  \\
       &\left.+  2 \, c_B \, \bra{D^+\pi^-} | \bar{d}_L \gamma_\mu t^a u_L 
\bar{c}_L \gamma^\mu t^a b_L  | \overline{B_d^0} \ket \right],
\label{BDpiFact}
\end{eqnarray}
and 
\begin{eqnarray}
 % \begin{split}
    \mathcal{M}_{D^0\pi^0} = 4 \, \frac{G_F}{\sqrt{2}} 
V_\textrm{cb}V^*_\textrm{ud} & 
\left[ \left(c_B+\frac{1}{N_\textrm{c}}c_A \right)
     \bra{D^0} | \bar{c}_L \gamma_\mu u_L | 0 \ket \bra{\pi^0} | \bar{d}_L
\gamma^\mu b_L  | \overline{B_d^0} \ket \right. \nonumber \\
       &\left.+  2 \, c_A \bra{D^0\pi^0} | \bar{d}_L 
\gamma_\mu t^a b_L \bar{c}_L \gamma^\mu t^a u_L | \overline{B_d^0} \ket
         \right] \, .
% \end{split}
\label{BDpiNonFact}
\end{eqnarray}
Here the  terms proportional to $2 c_A$ and $2 c_B$ with color
 matrices inside the matrix elements are
 the genuinely non-factorizable contributions. These will be estimated
 in section IV.

Since $c_A$ is slightly bigger than  one and  $c_B$ of order $- 0.4$,
 we refer to the coefficients 
 \begin{equation}
  c_{f} \, \equiv \, \left(c_A+\frac{1}{N_\textrm{c}}c_B\right) \, \simeq \,
  1   \qquad ; \quad
  c_{nf} \; \equiv \;  \left(c_B+\frac{1}{N_\textrm{c}}c_A \right) \, \simeq
  \, 0 \; \, ,
\end{equation}
as favorable ($c_f$) and non-favorable ($c_{nf}$) coefficients, respectively.
Thus, the decay mode
 $\overline{B_d^0} \rightarrow  D^+\pi^-$ has a sizeable factorized amplitude
proportional to $c_f$. In contrast, the decay mode
$\overline{B_d^0} \rightarrow D^0\pi^0$ has a factorized amplitude
proportional to the  non-favorable coefficient $c_{nf}$ which is close to
zero. In this case we expect the non-factorizable term (involving
color matrices) proportional to $2 c_A$ to be dominant,- i.e.  
the last line of eq. (\ref{BDpiNonFact}) dominates.
A substantial part 
 of this paper is dedicated to the calculation of this non-factorizable
contribution to the $\overline{B_d^0} \rightarrow D^0\pi^0$ decay
 amplitude, which in
 the factorized limit is proportional to the  non-favored coefficient $c_{nf}$.

\section{Perturbative QCD corrections to one loop within HQEFT}

Wilson coefficients for four quark operators for non-leptonic decays
have been first calculated at the one loop level \cite{QCDloop1}, and
later at the  two loop level \cite{QCDloop2}. In \cite{EHP,EFHP,MacDJoe}
the latter were used.
\begin{figure}[t]
\begin{center}
   \epsfig{file=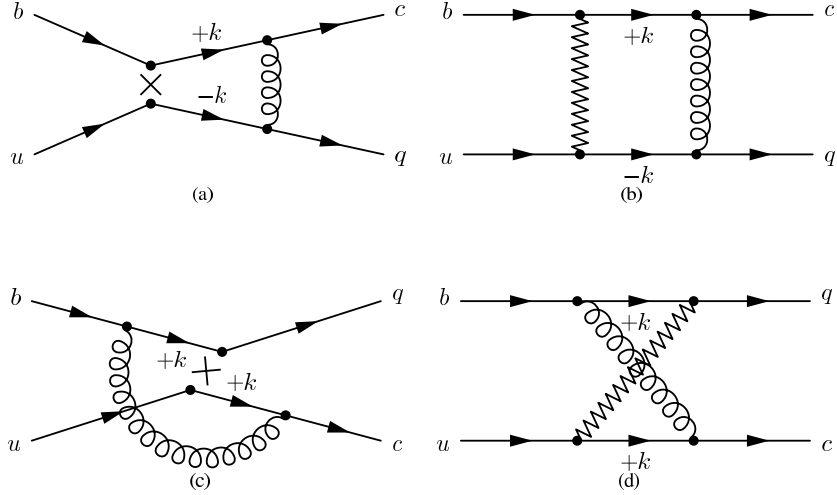,width=11cm}
\caption{QCD corrections for $Q_A$ (upper line) and $Q_B$ (lower line) when
 all quarks  are considered light, i.e. for $\mu > m_b$. In the left coloumn
 the weak interaction for an infinitely heavy $W$-boson is marked by a cross.
In the right coloumn, the weak interaction is marked by a zig-zag line.
(In the lower, right diagram, the zig-zag line represent a fictitious
 ``$W^0$'' exchange.) In all cases the curly lines represent gluon exchanges.}
\label{fig:BtoDpi}
\end{center}
\end{figure} 
 Here we will calculate the Wilson coefficient completely within
HQEFT at one loop level.
 Thus the heavy quarks will be described by the HQEFT Lagrangian
\cite{neu}: 
\begin{equation}
{\cal L}_\mathrm{HQET} = \bar{Q}_v \, (iv\cdot D) \, Q_v 
   + \mathcal{O}(1/m_Q) \; ,
\label{eq:LeffFINAL}
\end{equation}
where $Q_v$ is the reduced heavy quark field (often named $h_v$ in the 
literature), $v$ its four velocity and $m_Q$
the mass of the heavy quark.

As usual, the renormalization of the four quark operators are performed i 
several steps: First,  when the renormalization scale $\mu$  satisfies
$ m_b< \mu < M_W$, all the five quarks $b,c,s,d,u$ are considered light.
Then, for scales $m_c < \mu < m_b$, the $b$-quark is considered heavy
while the $c$-quark is still considered light.
Going further to the case
 $ \mu < m_c$, the $c$-quark is also considered heavy,
  Then the calculations are performed within strict
 HQEFT for both for the $b$ and the $c$ quark.
By assumption the various chiral quark models works below the chiral symmetry 
breaking scale $\Lambda_\chi \simeq \, $ 1 GeV. Also,  HL$\chi$PT
 is  applicable  below the scale $\Lambda_\chi$ \cite{EFHP,EHP,pider,BEF,epb}.
  Therefore we will match 
the perturbative calculations with our model at
  $ \mu = \Lambda_\chi $.
For renormalization scales $\mu$ in the region  $ m_b< \mu < M_W$,
where all the involved quarks are considered to be light, we obtain 
 the well known result
\begin{eqnarray}
c^{(0)}_A(\mu)& 
  = \frac{1}{2}\left[\left(\frac{\alpha_s(M_W)}{\alpha_s(\mu)}\right)^{6/23}
  + \left(\frac{\alpha_s(M_W)}{\alpha_s(\mu)}\right)^{-12/23}\right]
  \; , \\
c^{(0)}_B(\mu)& 
  = \frac{1}{2}\left[\left(\frac{\alpha_s(M_W)}{\alpha_s(\mu)}\right)^{6/23}
  - \left(\frac{\alpha_s(M_W)}{\alpha_s(\mu)}\right)^{-12/23}\right]
\; ,
\label{eq:c-scaled-once}
\end{eqnarray}
reflecting that the anomalous dimension matrix for the operator basis
$Q_\pm = (Q_B \pm Q_A)$ is diagonal.

For scales $\mu$ satisfying 
 $m_c < \mu < m_b$, the $b$-quark is considered to be heavy, while the
 $c$-quark is still light. In this range of $\mu$
 we find that some of the diagrams which contributed for 
 $ m_b< \mu < M_W$ are now zero. As a consequence, in the $(Q_A, Q_B)$
 basis the anomalous dimension matrix is now (using the definition 
$\gamma \equiv  (\alpha_s/2 \pi)  \, \hat{\gamma} $):
\begin{eqnarray}
\hat{\gamma} ( m_c< \mu < m_b)\; = \; \frac{1}{2} 
\left(\begin{array}{cc}
- 1 & 3 \\ 3 & - 1   
\end{array}\right)
\; \,  ,
\end{eqnarray}
which is half of what it is above $\mu = m_b$.
The beta function to lowest order is proportional to 
$b^{(1)}_0= 11- 2 \, N_f/3 \, ,$  where $N_f$ is the number of effective flavors.
With the bottom quark integrated out, $N_f = 4$, thus
$b^{(1)}_0 = 25/3$. Defining the quantity
\begin{eqnarray}
D(\mu) \equiv \left(\frac{\alpha_s(m_b)}{\alpha_s(\mu)}\right)^{3/25}
 \left(\frac{\alpha_s(M_W)}{\alpha_s(m_b)}\right)^{6/23} \; ,
\end{eqnarray}
we obtain for $m_c < \mu < m_b$ the Wilson coefficients:
\begin{eqnarray}
c^{(1)}_A(\mu)& 
  = \frac{1}{2}\left[ D(\mu) \; + \; \left(D(\mu)\right)^{-2} \right]
\; \, ,  
\nonumber \\
c^{(1)}_B(\mu)& 
 = \frac{1}{2}\left[ D(\mu) \; -  \; \left(D(\mu)\right)^{-2}
\right]
\; \, ,
\label{eq:c-scaled-twice}
\end{eqnarray}

\begin{figure}[t]
\begin{center}
   \epsfig{file=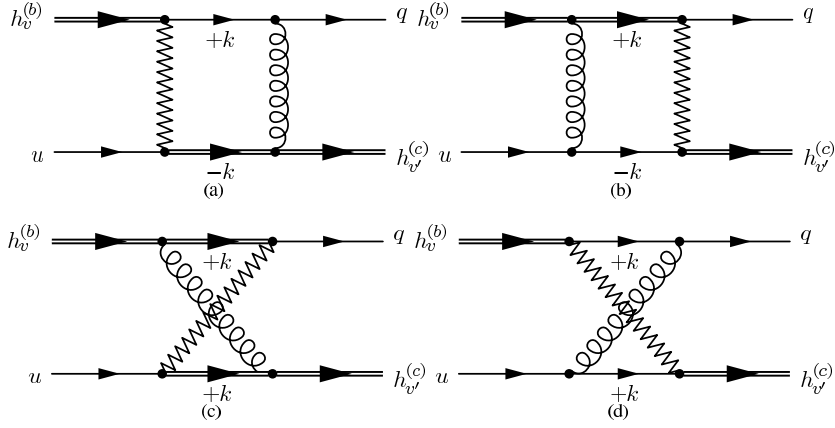,width=11cm}
\caption{QCD corrections for $Q_A$ (upper line) and $Q_B$ (lower line)
 in the case
 $\mu < m_c$, when both the $b$- and the $c$-quark are considered to be 
 heavy. The heavy quarks are represented by double lines. The zig-zag 
and curly lines have the same meaning as in FIG. 1.}
\label{fig:BtoDpiHL}
\end{center}
\end{figure} 

For  the range $ \Lambda_\chi < \mu < m_c$, where the $b$- and  the $c$-quark
are both considered as heavy,
 we obtain a more non-standard anomalous dimension matrix
\begin{eqnarray}
\gamma ( \Lambda_\chi < \mu < m_c) \, = \, \frac{1}{2} 
\left(1+\frac{2}{3}\omega \, r(\omega)\right)
\left(\begin{array}{cc}
0 & 0 \\ 3 & - 1 
\end{array}\right) \; \, .
\end{eqnarray}
 Then we finally get the result for $\Lambda_\chi < \mu < m_c$ the
 coefficients:
\begin{eqnarray}
c_A^{(2)}(\mu) \, = \, c_A^{(1)}(m_c) \qquad ; \qquad 
c_B^{(2)}(\mu) \, = 3 (1 - \tau) \, c_A^{(1)}(m_c) \, + \,
  \tau \, c_B^{(1)}(m_c)
\label{eq:cB-final}
\end{eqnarray}
where
\begin{equation}
\tau \, \equiv \,
 \left(\frac{\alpha_s(m_c)}{\alpha_s(\mu)}\right)^{\overline{\omega}}
\qquad ; \qquad 
\bar{\omega} \, \equiv \,  -\frac{1}{18}\left(1 \, + 
\, \frac{2}{3} \, \omega \, r(\omega) \right) \; ,
\end{equation}
the function $r(\omega)$ being the well known
\begin{equation}
r(\omega) \, \equiv \,  \frac{1}{\sqrt{\omega^2-1}}\left(\omega \, + 
\, \sqrt{\omega^2 -1} \right) \; .
\end{equation}
An analogous result for $b \rightarrow d c \overline{c}$ has been
obtained in \cite{GKMWF}.

We observe that  $c_A$ is not  further renormalized
 below $\mu = m_c$, while $c_B$ and thereby $c_{nf}$ get a small
 additional renormalization through the factor $\tau$ for 
$\Lambda_\chi =  \mu < m_c$.  Numerically,  $\tau$ is close to one.
At $\mu = \Lambda_\chi \simeq $ 1 GeV and the relevant value 
$\omega \simeq 1.6$, 
we have  $c^{(2)}_A \simeq 1.2$ and $c^{(2)}_B \simeq -0.44$, giving
 $c_f \simeq 1.0$ and  $c_{nf} \simeq -0.04$.

From the numerical point of view, the calculation performed in this
 section has not given us much new information. However, we think it is
 useful to have a calculation performed completely within HQEFT, and to
 our knowledge this is not presented anywhere else in the literature.

\section{An  energetic light quark  effective description (LEET$\delta$)}
\label{LEET}

An energetic light quark might, similarly to a heavy quark,  carry
 practically  all the energy $E$ of
the meson it is a part of (i.e. it has momentum fraction $x$ close to one).
 But   the mass of the energetic quark  is close to
 zero compared to $m_Q$ and $E$, which are assumed to be of the same 
order of magnitude. Assuming that the energetic light 
 quark is coming from the decay of a
 heavy quark $Q$ with momentum $p_Q =  m_Q \, v \, + k$, the momentum of
 the energetic quark $q$ can be written
\begin{equation}
p^\mu_q = E \, n^\mu + k^\mu \quad ,   
\qquad \;   |k^\mu| \ll |E \, n^\mu| \quad , \qquad  m_q \ll E \; ,
\label{eq:pq}
\end{equation}
where $m_q$ is the light quark mass and  $n$ is the light-like four
 vector which might be chosen to have 
the space part  along the z-axis,
 $n^\mu = (1; 0, 0, 1)$, 
 in the frame of the heavy quark where $v = (1;0,0,0)$.
Then  $(v\cdot n) = 1$ and  $n^2 \, = \, 0$.
Inserting this in the  regular quark propagator, we obtain
\begin{equation}
 S(p_q) \, = \, 
% \frac{1}{\gamma \cdot p_q - m_q} = 
\frac{\gamma \cdot p_q + m_q}{p_q^2 -m_q^2} = 
\frac{E \gamma \cdot n + \gamma \cdot k + m_q}{2 E n\cdot k + k^2-m_q^2}.
\end{equation}
In the limit where the approximations
in (\ref{eq:pq}) are valid, we obtain the propagator
\begin{equation}
 S(p_q) \; \rightarrow \;  \frac{\gamma \cdot n}{2 n \cdot k} \; \;  .
\label{eq:LEETprop}
\end{equation}
This propagator is the starting point for the Large Effective Theory
(LEET) constructed
in ref.\cite{charles}.

Unfortunately, the combination of LEET  with $\chi$QM will lead to
 infrared divergent loop integrals  for $n^2=0$ 
(see section \ref{HLET+LEET}). 
Therefore, in the following  we modify the formalism and instead  of  $n^2=0$,
 we use $n^2 = \delta^2$, with $\delta = \nu/E$
where $\nu \sim \Lambda_{QCD}$,  such that $\delta \ll 1$. 
In the following we derive a
modified LEET \cite{charles} where we  keep $\delta \neq 0$
with $\delta \ll 1$. We call this construction LEET$\delta$ and 
 define  the  {\it almost} light -like vectors
\begin{equation}
 n = (1,0,0,+\eta),\qquad ; \quad 
 \tilde{n} = (1,0,0,-\eta),
\end{equation}
where
$\eta = \sqrt{1-\delta^2} $. This means that
\begin{eqnarray}
n^\mu + \tilde{n}^\mu = 2v^\mu \; \,  ,\, n^2 = \tilde{n}^2 = 
\delta^2 \, , \, 
v\cdot n = v\cdot \tilde{n} = 1 \; \; , \, n \cdot  \tilde{n} \, = 2 -
\delta^2 \, .
\end{eqnarray}
Using the above equations, we choose the set of projection
 operators given by
\begin{equation}
  \mc{P}_+ = \frac{1}{N^2}\gamma \cdot n(\gamma \cdot
  \tilde{n}+\delta) \quad , \; 
  \mc{P}_- = \frac{1}{N^2}( \gamma \cdot \tilde{n}-\delta)
\gamma \cdot  n  \; \, ,
%\quad ; \; 
%  N    = \sqrt{2 \, n \cdot  \tilde{n}} \, .
 \label{eq:LEETprojectors}
\end{equation}
where  $N \, = \sqrt{2 \, n \cdot  \tilde{n}} \;
 =  \; 2 \, + {\cal O}(\delta^2)$ .
 We  factor out the main energy dependence, just as was analogously 
done  in HQEFT, and  define the projected reduced quark fields $q_{\pm}$
\cite{charles}:   
\begin{equation}
  q_\pm(x) = e^{iEn\cdot x}\mc{P}_\pm q(x) \quad  , \; \;  
  q(x) = e^{-iEn\cdot x}\left[q_+(x) + q_-(x)\right].
 \label{eq:LEETqpm}
\end{equation}
The adjoint fields are 
\begin{eqnarray}
  \bar{q}_\pm = q^\dag_\pm \gamma^0 
              = e^{-iEn\cdot x} \bar{q} \,  \overline{\mc{P}}_\pm 
\qquad ;  \; \; \, \overline{\mc{P}}_\pm \equiv \gamma^0 \mc{P}_\pm^\dag \gamma^0
\end{eqnarray}

Following the procedure of \cite{charles}, we eliminate $q_-$ and
 obtain for
 $q_+   \equiv q_n$ the effective Lagrangian:
\begin{eqnarray}
  {\cal L}_{LEET\delta} \, = \, 
 \bar{q}_n \left(\frac{\gamma \cdot  \tilde{n} + \delta}{N} \right)
(i n  \cdot D) q_n
  + \frac{1}{E}\bar{q}_n \, X \, q_n + \mathcal{O}(E^{-2}) \; ,
\label{eq:LEETdelta-lagrangian}
\end{eqnarray}
which (for $\delta = 0$) is  the first part of the SCET  Lagrangian.
Equation (\ref{eq:LEETdelta-lagrangian}) yields the LEET$\delta$
 quark propagator
\begin{equation}
S_n(k) \, = \, \mc{P}_+ \,  \left[\frac{\gamma \cdot \tilde{n} +
                   \delta}{N}(n \cdot k)\right]^{-1}
                   = \frac{\gamma \cdot n}{N(n\cdot k)} \; \; ,
\label{LEETpropd}
\end{equation}
which reduces to (\ref{eq:LEETprop}) in the  limit $\delta\rightarrow 0$.
In addition, for small $p_\perp^2$ \cite{SCET}, it coincides with
 the corresponding SCET-propagator.
 Our  $\mathcal{O}(E^{-1})$ term is given by
\begin{eqnarray}
X \, =  \, -\frac{1}{2}(i\gamma \cdot D) \gamma \cdot v
     \left[(i\gamma \cdot D) - \frac{(\gamma \cdot \tilde{n} +
  \delta)}{N}(i n \cdot D)\right] \nonumber \\ 
    -\frac{1}{2}\left[\gamma \cdot (i\stackrel{\leftarrow}{D}) 
     - \frac{(\gamma \cdot \tilde{n} + 
\delta)}{N}(i n\cdot \stackrel{\leftarrow}{D})\right]
      \gamma \cdot v (i \gamma \cdot D)  \; .
\end{eqnarray}

Based on LEET, it was found  \cite{charles}, 
in the formal limits $M_H \rightarrow \infty$ and 
$E\rightarrow\infty$, that a heavy $H-$ ( $B-$ or maybe also $D-$)
 meson decaying by a vector weak current $V^\mu$ to a light pseudoscalar
 meson $P$ has a matrix element
$\langle P \, | \, V^\mu \, | \, H \rangle$ of the form
\begin{equation}
\langle P| V^\mu |H \rangle = 2E\left[\zeta^{(v)}(M_H,E) \, n^\mu 
                      + \zeta^{(v)}_1(M_H,E) \, v^\mu  \right] \; , 
\label{HECurrent}
\end{equation}
where
\begin{equation}
\zeta^{(v)}   = C \frac{\sqrt{M_H}}{E^2} \quad , 
\; \, C \sim (\Lambda_\textrm{QCD})^{3/2} \qquad , \quad 
\frac{\zeta^{(v)}_1}{\zeta^{(v)}} \sim \frac{1}{E} \; \; .
\label{eq:charlesEq}
\end{equation}
This behavior is constistent  with the energetic quark
 having  $x$  close to one, where $x$ is the quark 
 momentum fraction  of the outgoing pion \cite{charles}.

\section{Extended chiral quark model for heavy and energetic light
  quarks (LE$\chi$QM)}
\label{HLET+LEET}

The  chiral quark model ($\chi$QM) \cite{chiqm,pider}  and the
 Heavy-Light Chiral Quark Model (HL$\chi$QM)
 \cite{ahjoe}, include meson-quark couplings and thereby
 allow us to calculate amplitudes and chiral  Lagrangians
 for processes involving heavy
 quarks and  low energy light
quarks. In this section we will extend these models to include also
 hard, energetic light quarks.

For the pure light sector the $\chi$QM Lagrangian  can be written 
as \cite{chiqm,BEF}:
\begin{equation}
 {\cal L}_{\chi QM} = \bar{\chi}\left( \gamma \cdot (i D + \mathcal{V})
       +\gamma \cdot \mathcal{A} \, \gamma_5 - m\right)  \chi  \; ,
\label{Light-Lagr}
\end{equation}
where $m$ is the  constituent mass term
being due to chiral symmetry breaking. (The small current mass 
term is neglected here).
 Here we have 
 introduced the flavor rotated fields $\chi_{L,R}$: 
\begin{eqnarray}
                 \chi_L =  \xi^\dagger \, q_L \qquad , \quad 
                 \chi_R =  \xi  q_R \; ,
\label{xirot}
\end{eqnarray}
where $q$ is the light quark flavor triplet and:
\begin{eqnarray}
            \xi \,= \; \exp\{i \Pi/f\}  \qquad , \quad 
\Pi &=& \left(\begin{array}{ccc}
     \frac{\pi^{0}}{\sqrt{2}} + \frac{\eta}{\sqrt{6}} 
     & \pi^{+} & K^{+} \\
     \pi^{-} & -\frac{\pi^{0}}{\sqrt{2}} +
     \frac{\eta}{\sqrt{6}} & K^{0} \\
     K^{-} & \bar{K}^{0} & 
     -\frac{2\eta}{\sqrt{6}}\end{array}\right) \; .     
\label{xidef}
\end{eqnarray}
Further, 
$\mathcal{V}_\mu$ and $\mathcal{A}_\mu$ are vector and axial vector fields,
 given by
\begin{equation}
 \mathcal{V}_\mu \equiv \frac{i}{2}(\xi^\dagger \partial_\mu\xi +
 \xi\partial_\mu\xi^\dag) \quad , \quad 
 \mathcal{A}_\mu \equiv -\frac{i}{2}(\xi^\dagger \partial_\mu\xi -
 \xi\partial_\mu\xi^\dag)  \; \; .
\end{equation}

To couple the heavy quarks to mesons there are additional meson-quark 
couplings within HL$\chi$QM \cite{ahjoe}:
\begin{equation}
{\cal L}_\mathrm{int} = -G_H \left[\bar{\chi}_a \, \bar{H}^a_v \, Q_v 
+ \bar{Q}_v \, H^a_v \, \chi_a\right]  \; ,
\end{equation}
where $a$ is a SU(3) flavor index,  $Q_v$ is the reduced heavy quark field,
 and  $H_v$ is the corresponding 
heavy $(0^-,1^-)$ meson field(s):
\begin{equation}
H_v = \mc{P}_+(v)\left(\gamma\cdot P^* - i\gamma_5 \, 
P_5\right) \; ,
\end{equation}
where $ \mc{P}_+(v) = (1 + \gamma \cdot v)/2$ is a projection operator.
Further, 
$P^*_\mu$ is  the  $1^-$ field  and $P_5$ the $0^-$ field.  
These  mesonic fields enter the Lagrangian of
 HL$\chi$PT: 
\begin{equation}
{\cal L}_{\mathrm{HL}\chi\mathrm{PT}}  \, = 
 \, - Tr(\bar{H_v} \, i v_\mu \partial^\mu H_v)
\,  + \, Tr(\bar{H_v}^a \, H_v^b \, v_\mu \, \mc{V}^\mu_{ba})  
 \, - \ga \, Tr(\bar{H_v}^a \, H_v^b \, \gamma_\mu \, \gamma_5 
\mc{A}^\mu_{ba}) \; \, ,  
\label{eq:L_HLchiPT}
\end{equation}
where $a,b$ are SU(3) flavor indices. The quark-meson coupling $G_H$  
is  determined within the HL$\chi$QM to be \cite{ahjoe} given by:
\begin{equation}
G_H^2 \; = \; \frac{2 m}{f_\pi^2} \, \rho \; \, ,
\label{GHcoupling}
\end{equation}
where $\rho$ is a hadronic quantity of order one.

For hard light quarks and chiral quarks coupling
to  a hard light meson multiplet field $M$, we extend the ideas of 
$\chi$QM and HL$\chi$QM, and assume that the energetic light mesons 
couple to light quarks with a derivative coupling to an axial current: 
\begin{eqnarray}
{\cal L}_{\mathrm{int}q}
 \; \sim \;  
\bar{q} \, \gamma_\mu \gamma_5(i \, \partial^\mu M) \, q \; \, .
\label{Ansatz}
\end{eqnarray}
We  combine  LEET$\delta$ with the $\chi$QM
and assume that the ingoing light quark and the outgoing meson are energetic,
 and we pull out a factor   $\exp{(\pm i E n \cdot x)}$ as in
 (\ref{eq:LEETqpm}).
To describe  (outgoing) light  energetic mesons, we use an octet 
$3\times 3$ matrix field $M =\exp{(+ i E n \cdot x)} \,  M_n \, $, 
where  $M_n$ has 
 the same form as $\Pi$ in (\ref{xidef}):
\begin{eqnarray}
M_n &=& \left(\begin{array}{ccc}
     \frac{\pi^{0}_n}{\sqrt{2}} + \frac{\eta_n}{\sqrt{6}} 
     & \pi^{+}_n & K^{+}_n \\
     \pi^{-}_n & -\frac{\pi^{0}_n}{\sqrt{2}} +
     \frac{\eta_n}{\sqrt{6}} & K^{0}_n \\
     K^{-}_n & \bar{K}^{0}_n & 
     -\frac{2\eta_n}{\sqrt{6}}\end{array}\right) \; \, .
\end{eqnarray}
Here $\pi^{0}_n$, $\pi^{+}_n$, $K^{+}_n$ etc. 
are the (reduced) meson fields  corresponding to  energetic light mesons
 with momentum $\sim E n^\mu$. 

Combining  (\ref{Ansatz}) with the use of the rotated
 soft quark fields in (\ref{xirot}) and  using 
$\partial ^\mu \rightarrow i E \,  n^\mu$
 we  arrive at the following  ansatz for the LE$\chi$QM interaction Lagrangian:
\begin{eqnarray}
{\cal L}_{\mathrm{int}q \delta}
  \; = \; 
         \, G_A \, E 
\bar{\chi} \,  (\gamma \cdot n) \, Z \, q_n \, + \, h.c.
\; \, , 
\label{eq:LEETdHLciQM}
\end{eqnarray}
where $q_n$ is the reduced field corresponding to an energetic light
 quark having momentum
fraction close to one (see \ref{eq:LEETdelta-lagrangian}),
  and $\chi$ represents a soft quark
 (see Eq. (\ref{xirot})). Further,  
$G_A$ is an unknown coupling to be determined later by physical requirements.
Further,
\begin{eqnarray}
Z         = \xi M_R \, R - \xi^\dag M_L \, L \; \, .
\end{eqnarray}
Here $M_L$ and $M_R$ are both equal to $M_n$, but they have formally
different transformation properties, 
This is analogous to the use of
quark mass matrices  ${\cal M}_q$ and ${\cal M}_q^\dagger$  in standard
 {\it Chiral Perturbation Theory} ($\chi$PT). They are in practise equal,
 but have formally different transformation properties.

The axial vector coupling introduces a $\gamma \cdot n$ factor to the vertex
(see (\ref{eq:LEETdHLciQM})), which  simplifies the Dirac algebra
 within the loop integrals.
\begin{figure}[t]
\begin{center}
   \epsfig{file=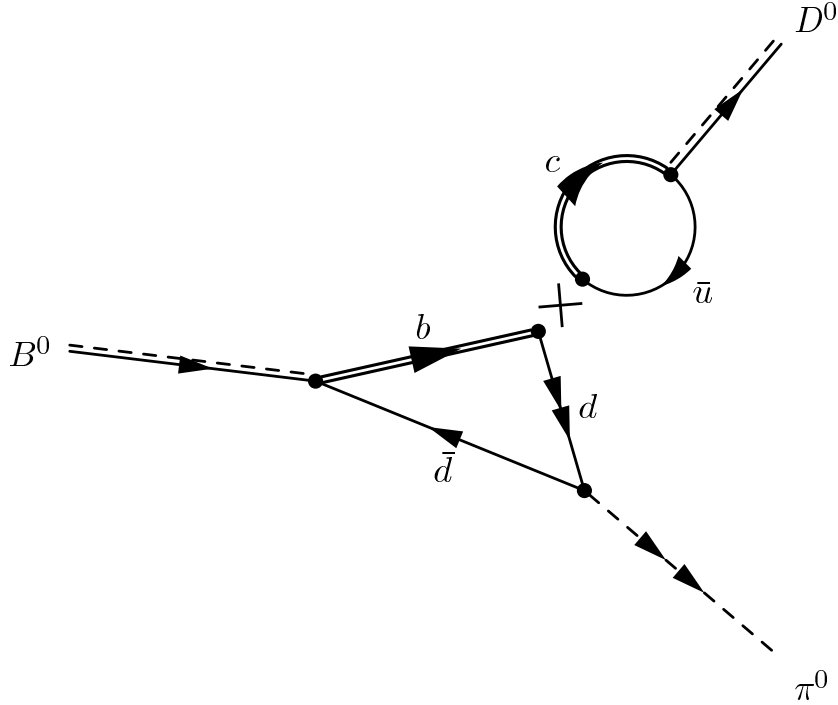,width=11cm}
\caption{The factorized contribution to the $B^0 \rightarrow D^0\pi^0$ decay,
         as described in combined  HL$\chi$QM and LE$\chi$QM. Double lines, 
single lines and the single line with two arrows are representing heavy
 quarks, light soft quarks and light energetic quarks, respectively. 
Heavy mesons are represented by a single line combined with a parallel
 dashed line, and the light energetic pion is represented by a dashed
 line with double arrow.}
\label{fig:BDpiF}
\end{center}
\end{figure} 
In order to calculate the non-factorizable contribution, we must
first find a value for the coupling  $G_A$ in (\ref{eq:LEETdHLciQM})
 assumed to bind a
large energy light quark and a soft (anti-) quark to an energetic light meson.
This will be done by requiring that our model should be constistent with
the equations (\ref{HECurrent}) and  (\ref{eq:charlesEq}). 
Applying the Feynman rules of LE$\chi$QM 
 we obtain the following bosonized current (before soft gluon emission forming 
the gluon condensate is taken into account):
\begin{eqnarray}
J_0^\mu (H_{v} \rightarrow M_n) =   -N_c \int \dbar k
  \textrm{Tr} \left\{ 
\gamma^\mu L 
\,i S_v(k) 
\left[-iG_H H_{v} \right]
\,i  S_\chi(k) \,  
\left[i E \, G_A \, \gamma \cdot n \, Z \right] \,i S_n(k) \, 
  \right\} \, ,
\label{eq:M0start}
\end{eqnarray}
where  $\dbar k \, \equiv \, d^d k/(2 \pi)^d $ ($d$ being the dimension of
 space-time),
 and 
\begin{eqnarray}
S_v(k) \, = \, \frac{P_+(v)}{v \cdot k} 
\quad , \; S_\chi(k) \, = \, 
\frac{(\gamma \cdot k + m)}{k^2 - m^2}
 \quad  , \, S_n(k) \, = \,
\frac{\gamma \cdot n}{N \, n \cdot k} \; \, ,
\label{propagators}
\end{eqnarray}
are the propagators for the reduced heavy quark fields ($Q_v$ in eq. 
(\ref{eq:LeffFINAL})), light constituent
 quarks ($\chi$ in eq.(\ref{Light-Lagr})), and the reduced light energetic 
quark fields  ($q_n$ in (\ref{eq:LEETdelta-lagrangian})), respectively. 
(Below we will use the leading order value $N=2$).

\vspace{0.2cm}
It should be emphasized that for the loop diagram for $B \rightarrow \pi$ 
in Fig. \ref{fig:BDpiF} (lower part of the diagram), we have the following
 picture: The large energy ($M_B \simeq m_b$) of the heavy $b$-quark 
and the large energy ($E \simeq M_B/2$) of the hard $d$-quark are floating
 through the (lower part of the) loop diagram. The loop momenta of
 the reduced quark fields for the  heavy quark, energetic  light quark are
 then carrying the  same soft loop momentum $k$ 
(with $|k| < \Lambda_\chi \, \simeq \; $ 1 GeV)
 as the soft light (anti-) quark 
($\overline{d}$), which  justifies the use of our model.     
\vspace{0.2cm}

The presence of the left projection operator $L$ in $Z$ ensures that we only
 get contributions from the left-handed part, 
 that is, 
$Z  \longrightarrow  -\xi^\dag M_L \, L$. 
The momentum integrals have the form
\begin{eqnarray}
K_{r s t} = \int  \, \frac{\dbar k}{(v \cdot k)^r \, (k \cdot n)^s \,
  (k^2 - m^2)^t} \; \, ,
\label{eq:Krst}
\end{eqnarray}
\begin{eqnarray}
K^\mu_{r s t} = \int  \,
 \frac{\dbar k \; k^\mu}{(v \cdot k)^r \, (k \cdot n)^s \, (k^2-m^2)^t}
  = K_{rst}^{(v)} v^\mu + K_{r s t}^{(n)} n^\mu \; \, , 
\label{eq:Kmu}
\end{eqnarray}
where $r,s,t$ are integer numbers.
These integrals have the  important property that
 $K_{r s t}^{(n)}$ dominates over 
 $K_{r s t}^{(v)}$
 and $K_{r s t}$ with one power of $1/\delta$. In the present model, we choose
$\nu = m$ which is of order $\Lambda_{QCD}$. Thus $\delta = m/E$ in the 
following.
 
The  contribution in (\ref{eq:M0start}), corresponding to the
 $B \rightarrow \pi$ part of Fig. \ref{fig:BDpiF},  with no gluon condensate
contribution included, contains $K_{111}$ and $ K_{111}^{\mu}$ and turns out to
be proportional to the formally linearly divergent  integral
$I_{3/2}$ \cite{ahjoe}.
 There are also other contributions with two emitted soft gluons 
 making a condensate \cite{pider,ahjoe}. To calculate emission of soft
 gluons we have used the framework of Novikov et al. \cite{Nov}. 
In this framework the ordinary vertex containing the gluon field
$A_\mu^a$ will be replaced by the soft-gluon version containing the
soft gluon  field tensor $G^a_{\mu \nu}$:
\begin{equation} 
 i g_s t^a \Gamma^\mu \, A_\mu^a \; \rightarrow  \; - \, \frac{1}{2}
 \, g_s \, t^a \, \Gamma^\mu \; G^a_{\mu \nu} 
\frac{\partial}{\partial p_\nu}\,  .... |_{p=0} \quad  ,
\label{Vertex}
\end{equation}
where $p$ is the momentum of the soft gluon. (Using this framework one has
 to be careful with the  momentum routing 
 because  the gauge where 
$x^\mu \, A^a_\mu =0$ has been used.)
 Here $\Gamma^\mu = \gamma^\mu \; , v^\mu $,
 or $n^\mu \, (\gamma \cdot \tilde{n} \, + \, \delta)/N$
for a light soft quark, heavy quark, or light energetic quark,
respectively. Our loop integrals are a priori depending on the 
gluon momenta $p_{1,2}$ which
are sitting in some propagators. These gluon momenta  disappear after
 having used the
procedure  in (\ref{Vertex}). It is understood that the
 derivatives in (\ref{Vertex}) 
have to be taken with 
respect to all propagators in the loop integral.

There is a contribution to the $H_{v} \rightarrow M_n$ current 
where two soft gluons are emitted from the
light quark line. This contribution contains  $K_{114}$ and $
K_{114}^{\mu}$ and is finite. Emission from the heavy quark or light
energetic quark are expected to be suppressed. This will be realized
in most cases because the gluon tensor is antisymmetric, and therefore such
contributions are proportional to
\begin{equation} 
G^a_{\mu \nu} v^\mu \, v^\nu \; = \; 0 \quad , \quad {\mbox or} \quad 
G^a_{\mu \nu} n^\mu \, n^\nu \; = \; 0 \; .
\label{SupprGcond}
\end{equation}
However, there are also contributions proportional to  :
\begin{equation} 
G^a_{\mu \nu} v^\mu \, n^\nu \; \neq \; 0  \quad , 
\label{NonSupprGcond}
\end{equation}
analogous to what happens in some diagrams for the 
Isgur-Wise diagram where there are two different 
velocities $v_b$ and $v_c$ \cite{KresJ}. In that case the corresponding
contributions are proportional to $(v_b \cdot v_c \, -1)$ which is zero for 
$v_c \rightarrow v_b$. 
Such  contributions (proportional to $K_{331}$ and $K_{331}^{\mu}$)
appear within our calculation when two soft
 gluons are emitted from the heavy quark line. (This statement is
 however gauge dependent. With another momentum routing such a
 contribution would come from another diagram. But summing all 
diagrams, gauge invariance is fulfilled).

 Using the prescription \cite{pider,BEF,ahjoe,Nov}
\begin{equation}
 g_s^2G^a_{\mu\nu}G^a_{\rho\lambda} \rightarrow 4\pi^2
\langle\frac{\alpha_s}{\pi}G^2\rangle                                    
\frac{1}{12}(g_{\mu\rho}g_{\nu\lambda}-g_{\mu\lambda}g_{\nu\rho}),
\label{GlueCorrel}
\end{equation}
for the gluon condensate one obtains a
 total bosonized current of the form
\begin{equation}
J_{tot}^\mu(H_v \, \rightarrow \, M_n) =
 - i \frac{G_H}{2} \, ( EG_A) \, \delta^2 \, 
   \textrm{Tr} \left\{ 
\gamma^\mu L 
H_v 
\left[ R^{(v)} + R^{(n)} \gamma \cdot n \right]
\xi^\dag M_L \right\} \; ,
\label{eq:Jmutot}
\end{equation}
where the relevant quantity needed is (to leading order in $\delta$):
\begin{eqnarray}
R^{(n)} = \frac{m}{\delta} \, F \quad ; \; \, F \, \equiv \; 
\frac{1}{m}\left( -i N_c I_{3/2} \; + \; 
\frac{\pi}{8 \cdot 16 m^3} [\frac{2}{3} \, - 1]
 \gc \right) \; .
\label{Rn}
\end{eqnarray}
Here the contribution $\sim 2/3$ within the parenthesis is
coming from   the
diagram where two gluons are emitted from the heavy quark line. This
contribution is due to (\ref{NonSupprGcond}).
Note that $F$ is dimensionless.

In order to obtain
the HL$\chi$PT Lagrangian terms
in (\ref{eq:L_HLchiPT}), 
 one calculates 
quark loops with attached heavy meson fields
and vector and axial vector fields $\mc{V}^\mu$ or $\mc{A}^\mu$. 
Then, as explained in previous papers \cite{chiqm,pider,BEF,ahjoe},
  logarithmic and linearly divergent integrals
$I_2$ and $I_{3/2}$ (as well as quadratic diverget integtral $I_1$)
will appear. These might be regularized, say, with ultraviolet cut-offs of 
order $\Lambda_\chi$ \cite{BHitE,BijRZAn}. 
 The explicite expressions of the divergent integrals 
in terms of the cut-offs will depend on the details of the regularization 
procedure. We will however not go into these details, but simply identify
 the divergent integrals by appropriate quantities regarded as physical
 within our model.
 That is, we the use   
 identification \cite{chiqm,pider} 
\begin{equation}
  - i  N_c I_2  \, = \, \frac{1}{4m^2} \left( f_\pi^2 \, - \,  
\frac{1}{24m^2}\langle\frac{\alpha_s}{\pi}G^2\rangle \right) \;
 \equiv \; \frac{f_\pi^2}{4m^2} \, \lambda \; \, ,
\label{I2}\end{equation}
for the logarithmically divergent integral, and \cite{ahjoe} 
\begin{equation}
-i N_c \, I_{3/2} \; = \; \frac{3 \, f_\pi^2}{8 m \, \rho} (1- g_A) \, + \,
 \frac{N_c m}{16 \pi}
 - \,  \frac{(8-3 \pi)}{256 \, m^3} \gc \; ,
%\langle\frac{\alpha_s}{\pi}G^2\rangle\right) 
%\, = \, g_\mc{A},
\label{eq:I32}
\end{equation}
for the linearly divergent integral.
The parameter $\lambda$ defined in (\ref{I2}) is of 
order $10^{-1}$ and rather sensitive to small variations in
 the parameters $m$ and $\gc$.
 Using (\ref{I2}) and (\ref{eq:I32}) can be shown that
the parameter $\rho$ in (\ref{GHcoupling}) is given by \cite{ahjoe,KresJ}
\begin{equation}
 \rho = \frac{(1+3g_A)}{4(1 + \frac{m^2N_c}{8 \pi f_\pi^2} \, - \,
 \frac{\eta_H}{2 m^2 \, f_\pi^2}  \langle\frac{\alpha_s}{\pi}G^2\rangle)} 
\; \, , 
\label{eq:rho}
\end{equation}
where   $\eta_H  = (8-\pi)/64$.
Then we obtain for the quantity $F$: 
 \begin{equation}
F \; = \; \frac{N_c }{16 \pi} \; + \; 
\frac{3 \, f_\pi^2}{8 m^2 \, \rho} (1- g_A)  \,
 - \,  \frac{(24 - 7 \pi)}{768 \, m^4} \, \gc \; \, .  
\label{FExpr}
\end{equation}
Numerically, $F \simeq 0.08$.

In order to fix $G_A$ in (\ref{eq:LEETdHLciQM}), we compare 
(\ref{HECurrent}) with (\ref{eq:Jmutot}).
In our  case where no  extra  soft  pions are going out,
  we put $\xi \rightarrow 1$, 
and $M_L \rightarrow k_M\sqrt{E}$, with the isospin factor
 $k_M = 1/\sqrt{2}$ for $\pi^0$ (while $k_M=1$ for charged pions).
 Moreover for the  $B$-meson
with spin-parity  $0^-$ we have
$ H_v 
\rightarrow P_+(v)(-i\gamma_5)\sqrt{M_H}$. Using this,
we obtain the traces
\begin{equation}
\textrm{Tr}\{\gamma^\mu L H_v\xi^\dag M_L\} 
  \rightarrow 
  -i\sqrt{M_H}(k_M\sqrt{E})v^\mu,
\label{eq:TrH1}
\end{equation}
\begin{equation}
\textrm{Tr}\{\gamma^\mu L H^{(+)}_v \gamma^\sigma \xi^\dag M_L\} 
  \rightarrow
   +i\sqrt{M_H}(k_M\sqrt{E})g^{\mu \sigma} .
\label{eq:TrH2}
\end{equation}
Then we obtain the following  matrix element of the current:
\begin{equation}
J_{tot}^\mu(H_v \, \rightarrow M_n) \, = \, 
 \frac{G_H}{2}(EG_A)\sqrt{M_H}(k_M\sqrt{E})\delta^2
  \left[ -R^{(v)}v^\mu + R^{(n)}n^\mu \; \; ,
  \right].
%\end{split}
\label{eq:MtotFTW}
\end{equation}
where
$R^{(v)}/R^{(n)} \,  \sim \,   \delta$, i. e. we obtain
 $R^{(v)}/R^{(n)}\rightarrow 0$ as
$E\rightarrow \infty$,
as we should according to (\ref{HECurrent}) and (\ref{eq:charlesEq}).
Using the equations (\ref{HECurrent}), (\ref{FExpr}), and 
(\ref{eq:MtotFTW}),  we obtain
 \begin{equation}
G_A \; = \; 
\frac{4 \zeta^{(v)}}{m^2 \, G_H \, F} \, \sqrt{\frac{E}{M_H}} \; ,
\label{GAExpr}
\end{equation}
where $\zeta^{(v)}$ is numerically known \cite{PBall} to be $\simeq \, 0.3$. 
 Within our model, the analogue of $\Lambda_{QCD}$ is the constituent
 light quark mass $m$.
To see the behavior of $G_A$ in terms of the energy $E$ we therefore 
 write $C$ in (\ref{eq:charlesEq}) as 
$C \, \equiv \, \hat{c} \, m^{\frac{3}{2}}$, and obtain
\begin{equation}
G_A \; = \;
 \left(\frac{4 \hat{c} f_\pi}{m \, F \, \sqrt{2 \rho}} \right)
\;  \frac{1}{E^{\frac{3}{2}}} \; \, ,
\label{GAExpr2}
\end{equation}
which explicitly displays the behavior $G_A \sim E^{-3/2}$. In terms of the
 number $N_c$ of colors, $f_\pi \sim \sqrt{N_c}$ and $F \sim N_c$ which gives
 the behavior $G_A \sim 1/\sqrt{N_c}$, i.e. the same behavior as for $G_H$.

\section{Non-Factorizable Processes in LE$\chi$QM}

\begin{figure}[t]
\begin{center}
   \epsfig{file=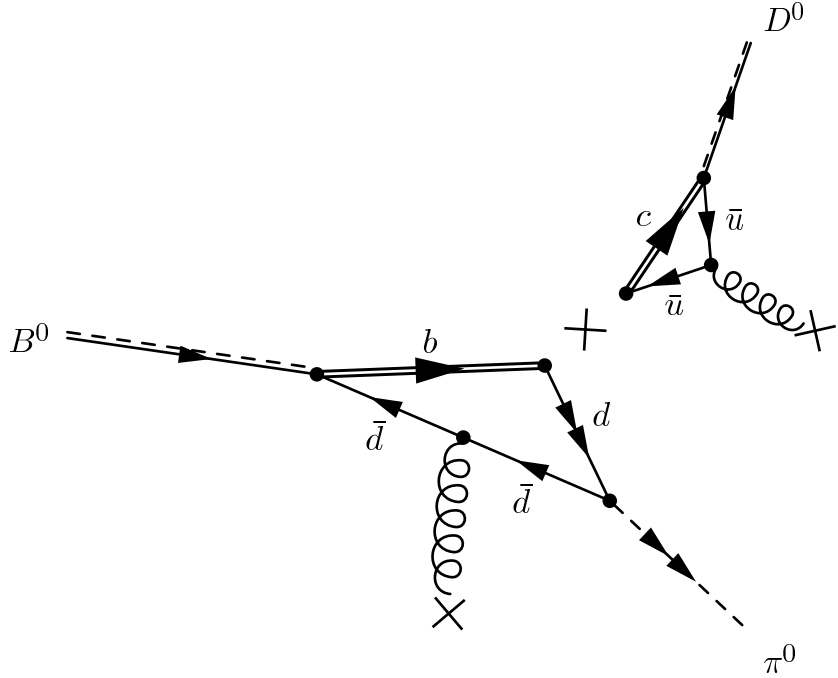,width=11cm}
\caption{The part of the non-factorizable contribution containing large energy
          light fermions and mesons. The curly lines represent soft gluons,
 and the cross in the end of these symbolizes the gluon condensate.
 Otherwise the various symbols have the same 
meaning as in FIG. 3}
\label{fig:BDpiNF}
\end{center}
\end{figure}

In this section we calculate the non-factorizable contribution to 
$\overline{B_d^0} \rightarrow \pi^0 D^0$ in eq. (\ref{BDpiNonFact}). This will be
formulated as a quasi-factorized product of two colored currents,
as illustrated in  Fig. \ref{fig:BDpiNF}.
Then the non-factorized aspects enters through color correlation
between the two parts, using eq. (\ref{GlueCorrel}).
Such a calculation within HL$\chi$QM is done previously \cite{EFHP}
for $\overline{B_{s,d}^0} \rightarrow  D^0 \overline{D^0}$, where the 
relevant colored
current for decay of a $D$-meson was calculated. What we will
calculate here is the colored current for  $B \rightarrow \pi$ with soft
 one gluon emission, within
the LE$\chi$QM presented in the preceding section;
 see diagram \ref{fig:BDpiNF}.

 Using the values for $G_A$ and $G_H$  from the preceding section, we 
find an expression  for the non-factorizable 
$\overline{B_d^0} \rightarrow D^0\pi^0$ decay amplitude, which
may be compared with experiment.

For a low energy quark interacting with one soft gluon, one might in
 simple cases use 
 the effective propagator \cite{RRY,BEF}
\begin{equation}
 S^G_1(k) \, = \, \frac{g_s}{4} t^a G^a_{\mu\nu}
\frac{(2 m \sigma^{\mu\nu}  \, + \left\{\sigma^{\mu\nu} , \gamma \cdot
  k \right\} )}{(k^2-m^2)^2} \; \, ,
\end{equation}
where $\left\{a,b\right\} \equiv a b + b a$ denotes the anticommutator.
This expression is consistent with the prescription in
(\ref{Vertex}),
and can be used for diagram 4.

Then we obtain the following contribution to the bosonized colored $B
\rightarrow \pi$ current corresponding to  diagram \ref{fig:BDpiNF} :
\begin{eqnarray}
  J^\mu_{1G}(H_v \, \rightarrow \, M_n)^a_4 \,  = \,  
    -\int \dbar k  \mathrm{Tr} \left\{ \gamma^\mu L t^a 
\, i S_v(k) \,                          
\left[-iG_H H_v \right]
\, i S_1^G(k) \, 
                        \left[i \, EG_A \, \gamma \cdot n Z \right]
\, i S_n(k) \; 
\right\} \, ,
 \label{eq:M1}
\end{eqnarray}

Once more, we deal with the momentum integrals of the types
in (\ref{eq:Krst})  and (\ref{eq:Kmu}).
 Taking the color trace, we obtain a
 contribution of the form
\begin{eqnarray}
  J^\mu_{1G}(H_{v} \rightarrow M_n)^a \, = \,
        g_s \, G^a_{\alpha \beta}  T^{\mu; \alpha \beta}(H_v \rightarrow M_n) \; ,
\label{eq:JtoM}
\end{eqnarray}
where the contribution from the  diagram \ref{fig:BDpiNF}
alone is  to leading order in $\delta$
\begin{eqnarray}
  T^{\mu; \alpha \beta}(H_{v} \rightarrow M_n)_4 \, = \,
          \frac{G_H \, G_A}{128 \pi}
 \epsilon^{\sigma \alpha \beta \lambda} \, n_\sigma
\mathrm{Tr} \left( \gamma^\mu L H_v
  \gamma_\lambda \,
    \xi^\dag M_L \right) \; ,
\label{eq:HtoM4}
\end{eqnarray}
where $E \cdot \delta = m$ has been explicitly used.

There is also a diagram 5 not shown where the soft gluon is emitted from
the energetic quark. This diagram is zero due to (\ref{SupprGcond}).
Furthermore, there is a diagram 6 not shown where the gluon is emitted from
the heavy quark which contains a non-zero part due to
(\ref{NonSupprGcond}). 
 In this case  we have to stick to the general rule in 
(\ref{Vertex}).
This gives the additional contribution 
\begin{eqnarray}
  T^{\mu; \alpha \beta}(H_{v} \rightarrow M_n)_6 \, = \,  i \, 
  \frac{G_H \, G_A }{64 \pi} v_b^\alpha \, n^\beta 
\mathrm{Tr} \left( \gamma^\mu L H_v
  \gamma \cdot n \, 
    \xi^\dag M_L \right)
\label{eq:HtoM6}
\end{eqnarray}
The total contribution to (\ref{eq:JtoM}) is given by the right hand
sides of (\ref{eq:HtoM4}) and  (\ref{eq:HtoM6}).
Below we will use all the expressions for the various  
$J^\mu_{1G}(H_{v} \rightarrow M_n)^a$ for a decaying $B$-meson,- i.e. we 
have $v= v_b$.

The colored $D^0$ current was found in \cite{EFHP} to be
\begin{equation}
  (\overline{Q^{(+)}_{v_C}} t^a \, \gamma^\alpha \,  q_L)_{1G} \rightarrow \,
 J^\mu_{1G}(\overline{H_{v_c}})^a
\, = \,   g_s G^a_{\alpha \beta} \; 
 T^{\mu; \alpha \beta}(\overline{H_{v_c}}) \; \, , 
\label{eq:D0cur}
\end{equation}
where 
\begin{equation}
    T^{\mu; \alpha \beta}(\overline{H_{v_c}}) \, = \, 
    -\frac{G_H}{64\pi}\mathrm{Tr}\left[\xi \gamma^\mu L 
     \left(\sigma^{\alpha \beta}- \frac{2\pi f_\pi^2}{m^2 N_c} \, \lambda \, 
\left\{\sigma^{\alpha \beta} \, , \,  \gamma \cdot v_c \right\}
\right) \overline{H_{v_c}} \right] \; \, ,
\label{eq:D0}
\end{equation}
where $\lambda$ is  defined in (\ref{I2}).

Now we use (\ref{GlueCorrel})
and also include the Fermi coupling
the Cabibbo-Kobayashi-Maskawa matrix elements, and the coefficient
 $2 c_A$ for the non-factorizable
contributions to the amplitude, where $c_A$ is the Wilson coefficient
 for the $\mc{O}_A$ local operator. Then 
we find an effective Lagrangian at mesonic level  relevant  for the 
non-factorizable contribution to $\bar{B}^0\rightarrow D^0\pi^0$:
\begin{eqnarray}
  \mc{L}^{LE\chi QM}_{\textrm{Non.fact.}} \, = \, 
    \frac{4 \pi^2  c_A}{3}\left(4\frac{G_F}{\sqrt{2}}V_{cb}V^*_{ud}\right)
\gc \, S(H_{v_b} \rightarrow M_n \, H_{v_c}) \; ,
\label{eq:effLagrMes}
\end{eqnarray}
where $S(H_{v_b} \rightarrow M_n \, H_{v_c})$ is the tensor product 
\begin{eqnarray}
S(H_{v_b} \rightarrow M_n \, H_{v_c}) \; \equiv \; 
 T^{\mu; \alpha \beta}(H_{v_b} \rightarrow M_n) \;
  T_{\mu; \alpha \beta}(\overline{H_{v_c}}) \; .
\label{eq:TensProd}
\end{eqnarray}

The four vector products $(v_b\cdot v_c)$, $(v_b\cdot n)$, and $(v_c\cdot n)$ 
can be related to physical parameters by the equations for
momentum and energy conversation. From
\begin{equation}
 M_Bv_b^\mu = M_Dv_c^\mu \, + \, E n^\mu \, \quad  , \quad 
\, E = \frac{M^2_B-M^2_D}{2M_B} \; \, ,
\end{equation}
we obtain (up to $\mathcal{O}(\delta^2)$)
\begin{eqnarray}
(v_b\cdot v_c) = \frac{M_B^2+M_D^2}{2M_BM_D} \quad , \; 
(v_b\cdot n) = 1 \quad \; , \quad
(v_c\cdot n) = \frac{M_B}{M_D} \; \, .
\label{eq:vbn2}
\end{eqnarray}
Using $\delta = m/E$ and  
(\ref{eq:vbn2}),  we find
 an explicite expression for $S(H_{v_b} \rightarrow M_n \, H_{v_c})$ in the case 
$\overline{B^0} \rightarrow D° \, \pi^0$: 
\begin{eqnarray}
S(\overline{B^0} \rightarrow \pi^0 D^0) \; = \; 
   \frac{G_H^2 \,  G_A }{32 \cdot 64 \pi^2 }
 \,   \sqrt{M_B \, M_D \, E}                  
  \left(\frac{M_B}{M_D}\right)  \,
\left(1 \; + \; \frac{6 \pi f_\pi^2}{N_c m^2} \, \lambda \right) 
\frac{1}{\sqrt{2}} \; \, .
  \; 
\label{ExplicS}
\end{eqnarray}
Inserting the expressions for $G_H$ in (\ref{GHcoupling}) 
and $G_A$ in (\ref{GAExpr}), we obtain
\begin{eqnarray}
S(\overline{B^0} \rightarrow \pi^0 D^0) \; = \; 
   \frac{\sqrt{2 \rho} \; \zeta^{(v)}}{8 \cdot 64 \pi^2 \, F \, f_\pi}
 \,   \sqrt{\frac{m}{ M_D}}                  
  \left(\frac{E \, M_B}{m^2}\right)  \,
\left(1 \; + \; \frac{6 \pi f_\pi^2}{N_c m^2} \, \lambda  \right) 
\frac{1}{\sqrt{2}} \; .
 \label{ExplicSE}
\end{eqnarray}

We will now compare this non-factorizable  amplitude for 
$\overline{B^0} \rightarrow D^0 \pi^0$ with the factorized amplitude which 
dominates $\overline{B^0} \rightarrow D^+ \pi^-$ :
\begin{eqnarray}
   \mathcal{M}_{D^+\pi^-} \, =
\, 4 \, \frac{G_F}{\sqrt{2}}V_\textrm{cb}V^*_\textrm{ud} \,
     \left(c_A+\frac{1}{N_\textrm{c}}c_B \right) \cdot
\left(\frac{1}{2} \, f_\pi \, E \, n_\mu \right) \cdot 
\left(\frac{1}{2}  \, \sqrt{M_B M_D} \, (v_b + v_c)^\mu \, \xi(\omega)\right)
\, ,
\label{BDpiFactAmp}
\end{eqnarray}
where  $\xi(\omega)$ is the Isgur-Wise function.

The ratio between the non-factorized and factorized amplitudes are now
\begin{eqnarray}
r \; = \;
 \frac{{\cal M}(\overline{B^0} \rightarrow \pi^0 D^0)_{\mbox{Non-Fact}}}
{{\cal M}(\overline{B^0} \rightarrow \pi^- D^+)_{\mbox{Fact}}} \; = \; 
\frac{c_A}{c_f} \, \frac{h}{(1 + \frac{M_D}{M_B})}
 \, \frac{ \zeta^{(v)}}{\xi(\omega)} \, \sqrt{\frac{m}{M_B}}
\label{ratio}
\end{eqnarray}
where $h$ is our model-dependent hadronic factor 
\begin{eqnarray}
h  \; = \; 
   \frac{\sqrt{\rho} \, \gc}{96 \cdot  \, F \, f_\pi^2 \, m^2}
\left(1 \; + \; \frac{6 \pi f_\pi^2}{N_c m^2} \, \lambda \right)  \; ,
  \; 
\label{ExplicSEr}
\end{eqnarray}
which  behaves as $h \sim 1/N_c$ with respect to color.

It will  be interesting how the ratio $r$ scales in the limit 
$M_B^2 \gg M_D^2 \gg m^2$. Then we use the scaling of  $\zeta^{(v)}$ given in 
 (\ref{eq:charlesEq}) with $C = \hat{c} \, m^{3/2}$ as in (\ref{GAExpr2}). 
The  scaling of $\xi(\omega)$ for $M_B^2 \gg M_D^2 $ is not so well 
established.  Under certain
 assumptions \cite{Oliver} it is found that the IW function $\xi(\omega)$
has the form
  \begin{eqnarray}
\xi(\omega) \; = \;
 \left(\frac{2}{1+ \omega}\right)^\gamma
\label{IWForm}
\end{eqnarray}
where $\omega = v_b \cdot v_c$.
In the  so-called BPS-limit\cite{Uraltsev} 
one obtains \cite{Oliver} $\gamma = 3/2$. 
The IW function calculated within a bag model \cite{HoZa} 
has the same form as in
(\ref{IWForm}). Within chiral HL$\chi$QM calculations, the IW function
 will have terms of the type in (\ref{IWForm}) for $\gamma=1$, and some 
terms which for $\omega \gg 1$ scale as $\ln \omega/\omega$ 
\cite{BHitE,HoZa,ahjoe,KresJ}, where $\omega \sim (M_B/2 M_D)$. 

Using the simple form (\ref{IWForm})  and (\ref{eq:vbn2}), we find for
 of $r$ for $M_B^2 \gg M_D^2 \gg m^2$:
\begin{eqnarray}
r \; \simeq \;  
\frac{c_A}{ c_f} \, \frac{h \,  \hat{c}}{4^{(\gamma-1)}} \, 
\frac{m^2}{(M_D)^{\gamma} \, (M_B)^{(2-\gamma)}}
\end{eqnarray}
 Anyway, our calculations show that the ratio $r$ of the amplitudes are 
suppressed by $1/N_c$ and by inverse powers of heavy meson masses, as expected.

Concerning numerical predictions from our model, we have to 
stick to eq. (\ref{ratio}). 
 The measured branching ratios for 
  $\overline{B_{d}^0} \rightarrow \, \pi^- D^{+}$
 and   $\overline{B_d^0} \rightarrow \pi^0 D^0$  are 
 $ \simeq \, (2.68 \pm 0.13) \times 10^{-3}$ and 
$(2.62 \pm 0.24) \times 10 ^{-4}$, 
respectively \cite{PDG}. In order to predict the experimental value solely 
with the mechanism considered in this section, we should have $r \simeq 1/3$. 
 For typical values 
$m \, \sim \, $ 200 to 220 MeV and $\gc \, \sim \, $ 300 to 320 MeV,
 we find that $h \, \sim \, $  3. Further, numerically
 $\zeta^{(v)} \, \simeq 1/3$ \cite{PBall},
and $\xi(\omega) \, \sim \, 2/3$,  and $(1 + M_D/M_B) \simeq 4/3$.  
 Thus we obtain a
 ratio $r$ of order 1/4, i.e. our model can account for roughly 3/4
 of the amplitude needed to explain the experimental branching ratio.

\section{Conclusion}

 We have presented perturbative QCD corrections for the quark process 
 $b\rightarrow  c\bar{u}q$ calculated completely
 within HQEFT at one loop level, and scaled with RGE down to $\mu =
 \Lambda_\chi \simeq \, $ 1 GeV. We have shown that the factorized 
amplitude for process
  $\overline{B_{d}^0} \rightarrow \,
\pi^0 \overline{D^{0}} \, $ is proportional to a Wilson
 coefficient combination close to zero. Thus the non-factorizable
 contribution dominate the amplitude for this decay mode.
To handle the non-factorizable contributions we have extended previous
 chiral quark models for the pure light quark case\cite{chiqm} used
 in \cite{pider,epb,BEF},  and the
 heavy light case\cite{ahjoe} used in
 \cite{ahjoeB,EHP,EFHP,EHbeta,MacDJoe,EHFiz},
 to include also energetic light quarks.
Thus,  within our framework, the heavy  and the energetic light quarks  
are represented by reduced  fields corresponding to the redundant 
soft(er) interactions obtained when we split off the hard momenta, being 
of order $m_b$ or $m_c$ 
for heavy quarks and $E \simeq m_b/2$ for the light energetic quark.

We have found that within our model we can account for 3/4 of the 
amplitude needed to explain the experimental branching ratio \cite{PDG}.
 In addition to our contributions one might think of mesonic loops  
like for  processes  of the type $B \rightarrow \,
D \, \overline{D} \, $ \cite{EFHP} and  $B \rightarrow \,
\gamma \, D \, $ \cite{MacDJoe}, but for such mesonic loops
one  has to inserted ad hoc farm factors, or they should be handled
 within dispersion relation techniques. In both cases such calculations
 are beyond   the scope of this paper.
Anyway, final state interactions should be present\cite{KaidVysot}.

{\acknowledgments}

JOE is supported in part by the Norwegian
 research council
 and  by the European Commision RTN
network, Contract No. MRTN-CT-2006-035482 (FLAVIAnet).
He also thanks T.B. Ness for useful comments.

\bibliographystyle{unsrt}

\begin{thebibliography}{99}


\bibitem{BBNS}
M. Beneke, G. Buchalla, M. Neubert, C. T. Sachrajda,  Phys. Rev.
Lett. {\bf 83}, 1914 (1999).

\bibitem{SCET}
 C.~W.~Bauer, S. Fleming, and M. Luke,
  Phys. Rev. D {\bf 63}, 014006 (2000);

 C.~W.~Bauer, S. Fleming, D. Pirjol, and  I. W. Stewart,
 Phys. Rev. D {\bf 63}, 114020 (2001); 


 C.~W.~Bauer, D. Pirjol, I. Z. Rothstein,
I. W. Stewart,  Phys. Rev. D {\bf 70}, 054015 (2004). 




\bibitem{EFHP} 
  J.O. Eeg, S. Fajfer , and  A. Hiorth,
  Phys.Lett. {\bf B570}, 46-52 (2003);\\
  J. O. Eeg,  S. Fajfer, and  A. Prapotnik
 Eur. Phys. J. {\bf C42},   29-36 (2005).\\
See also: J.O. Eeg, S. Fajfer, J. Zupan,  Phys. Rev.  {\bf D 64}, 
  034010 (2001).  

\bibitem{pQCD}
Y.~Y.~Keum, H-n. Li, A.I. Sanda, Phys.Lett. {\bf B504}, 6-14 (2001).



\bibitem{pQCDBDD}
Ying Li, Ciai-Diam L\"u, and Zhen-Jun Xiao,
 J. Phys.  {\bf G 31}, 273 (1999).


\bibitem{neu}See for instance
 M.~Neubert,  Phys. Rep. {\bf 245},  259 (1994).

A. V. Manohar and M.B. Wise: {\em heavy Quark Physics}, Cambridge
University Press, 2000.


\bibitem{itchpt} R.~Casalbuoni, A.~Deandrea, N.~Di~Bartolomeo, 
R.~Gatto, F.~Feruglio and  G.~Nardulli, \\
  Phys. Rep. {\bf 281}, 145 (1997).

\bibitem{IW}
C.G. Boyd and B. Grinstein,  
 Nucl. Phys. {\bf B451}, 177 (1995).


\bibitem{ahjoeB}
 A.~Hiorth and J.~O.~Eeg, 
 Eur. Phys. J. direct {\bf C30},  006 (2003), and references therein.


\bibitem{GriLe}
B.Grinstein and R.F. Lebed,  Phys.Rev. {\bf D60},  031302(R) (1999).



\bibitem{EHP}
 J.~O.~Eeg, A.~Hiorth, A.~D.~Polosa, 
 Phys. Rev. {\bf D 65}, 054030 (2002). 


\bibitem{MacDJoe}
J.A. Macdonald S\o rensen and J.O. Eeg, Phys. Rev. {\bf D 75}, 034015 (2007).
% hep-ph/0605078.


\bibitem{neupet}
M. Neubert and A. Petrov,  {\it Phys. Lett .} {\bf B 519},  50-56 (2001).

\bibitem{GD}
M.J. Dougan and B. Grinstein, Phys. Lett. {\bf B 252}, 583  (1991).


\bibitem{charles}
J. Charles, A. Le Yaouanc, L. Oliver, O. P\`ene and J.C. Raynal,\\
 {\it Phys. Rev.} {\bf D60},  014001 (1999).



\bibitem{Uglea}
C. Balzereit, T. Mannel, and W. Kilian, Phys. Rev. {\bf D 58}, 114029
(1998);

 U.~Agleatti, G:~Corb\`o, and L.~Trentadue , 
 {\it Int. J. Mod. Phys.} {\bf A 14},  1769 (1999);\\
 U.~Agleatti and G:~Corb\`o, 
 {\it Int. J. Mod. Phys.} {\bf A 15},  363 (2000).


\bibitem{chiqm}
See for example:
A.~Manohar and H.~Georgi,  Nucl. Phys.  {\bf B234},  189 (1984); \\
D.~Espriu, E.~de~Rafael and J.~Taron,  Nucl. Phys. {\bf B345}, 22
(1990);
and references therein.

\bibitem{pider}
A.~Pich and E.~de Rafael, Nucl. Phys. {\bf B358}, 311 (1991).

\bibitem{BHitE}
W.~A.~Bardeen and C.~T.~Hill,  Phys. Rev. {\bf D49}, 409 (1994);\\
A.~Deandrea, N.~Di~Bartolomeo, R.~Gatto, G.~Nardulli, and A.~D.~Polosa, \\
 Phys. Rev. {\bf D58},  034004 (1998);
  A.~Polosa,  Riv. Nuovo Cim. {\bf 23 N11},  1 (2000);\\  
D.~Ebert, T.~Feldmann  R.~Friedrich and H.~Reinhardt,
 Nucl. Phys. {\bf B434}, 619 (1995).

\bibitem{BijRZAn}
A.A. Andrianov and V.A. Andrianov,  Z. Phys. {\bf C 55} 435 (1992);\\
J. Bijnens, E. de Rafael, and H. Zheng, Z. Phys. {\bf C 62} 437 (1994).

\bibitem{LELeg}
Lars E. Leganger, Master Thesis, Univ of Oslo, 2007.


\bibitem{EHFiz}
J.O. Eeg and A. Hiorth, 
{\it Fizika} {\bf B14},  41-74 (2005)
[e-Print: hep-ph/0411393].


\bibitem{ahjoe} A.~Hiorth and J.~O.~Eeg,  Phys. Rev. {\bf D 66},
 074001 (2002).




\bibitem{BEF}
S.~Bertolini, J.O.~Eeg and M.~Fabbrichesi, 
 Nucl. Phys. {\bf B449}, 197 (1995); \\
V.~Antonelli, S.~Bertolini, J.O.~Eeg,
M.~Fabbrichesi and E.I.~Lashin, \\
 Nucl. Phys. {\bf B469},  143 (1996);
S.~Bertolini, J.O.~Eeg,
M.~Fabbrichesi and E.I.~Lashin,\\
  Nucl. Phys. {\bf B514},  63 (1998);
{\it ibid} {\bf B514}, 93 (1998).


%\bibitem{SoftGlu}
%Art on eff soft gluon prop

\bibitem{GrinPir}
B. Grinstein and  D. Pirjol,
{\it Phys.Lett.} {\bf B615},  213-220 (2005)
[hep-ph/0501237]. 


\bibitem{QCDloop1}
M.K. Gaillard and B. W. Lee, {\it Phys. Rev. Lett.} {\bf 33},  108 (1974); 

G. Altarelli and L. Maiani, {\it  Phys.Lett.} {\bf 52 B},  351 (1974);

A.I. Vainstein, V.I. Zakharov, L.B, and M. A. Shifman,
 {\it Zh. Exp. Theor. Fiz.}  {\bf 72}  1275 (1977) [{\it Sov. Phys. JETP}  45
 670-682 (1977)].


\bibitem{QCDloop2}
G. Buchalla, A. J. Buras and M. E. Lautenbacher 
%{\em Weak decays beyond leading logarithms}, 
Rev. Mod. Phys. {\bf 68},  1125  (1996),
[hep-ph/9512380], and references therein.

M. Ciuchini, E. Franco, G. Martinelli, and  L. Reina,
Nucl.Phys. {\bf B415},  403 (1994)
[hep-ph/9304257].



%\bibitem{BuBuLa}
%G. Buchalla, A. J. Buras, M. E. Lautenbacher  Rev. Mod. Phys.
%{\bf 68},  1125 (1996).

\bibitem{GKMWF}
B. Grinstein, W.~Kilian, T.~Mannel, and M.B. Wise,  Nucl. Phys. {\bf B363} ,
19 (1991);\\
R. Fleischer,  Nucl. Phys. {\bf B 412},  201 (1994).




\bibitem{epb}
J.~O.~Eeg and I.~Picek, Phys. Lett. {\bf B301}, 423 (1993); 
{\it ibid.} {\bf B323} 193 (1994); \\
A.E.~Bergan and J.O.~Eeg,  Phys. Lett. {\bf 390},  420 (1997).
 

\bibitem{Nov}
V. Novikov, A.I. Vainstein, V.I. Zakharov, L.B, and M. A. Shifman,\\
 {\it Fortschr. Phys.}  {\bf 32} No 11,   585-622 (1984).

\bibitem{PBall}
See for example:
P. Ball, JHEP {\bf 9809}:005 (1998) ;\\
F. De Fazio, T. Feldman,and T. Hurth, JHEP {\bf 0802}:031 (2008).

\bibitem{PDG}
Particle data Group, Phys. Lett. {\bf B 667}, 1 (2008).

\bibitem{Oliver}
F. Jugeau, A. Le Yaouanc, L. Oliver and J.-C. Raynal,
Phys. Rev. {\bf D 74},  094012 (2006).

\bibitem{Uraltsev}
N. Uraltsev, Phys. Lett. {\bf B 585},  253  (2004).

\bibitem{KresJ}
K.Kumericki and J.O. Eeg, arXiv:0910.5428 (hep-ph).

\bibitem{RRY}
L.J. Reinders, H. Rubinstein and S. Yazaki, Phys. Rep. {\bf 127}, 1 (1985).

\bibitem{EHbeta}
 A.~Hiorth and J.~O.~Eeg, 
 Eur.Phys.J.direct {\bf C39 S1}, 27-35 (2005) [hep-ph/0304247].

\bibitem{HoZa}
H. Hogaasen and M. Sadzikowski, Z. Phys. {\bf C 64}, 
 427 (1994)[hep-ph/9402279].

\bibitem{KaidVysot}
A.B.Kaidalov and M.I.Vysotsky, 
Phys. Atom. Nucl. {\bf 70}, 712-721 (2007)
[hep-ph/0603013].

\end{thebibliography}

\end{document}